\documentclass[eqsecnum,aps,superscriptaddress,11pt ]{revtex4}
\usepackage{latexsym,amsmath,amssymb,amsfonts}
\usepackage[a4paper, tmargin=2.54cm, bmargin=2.54cm, lmargin=2.54cm, rmargin=2.54cm]{geometry}
\usepackage{graphicx}
\usepackage{setspace}

\begin{document}

\title{{\it Ab initio} studies of electron correlation and Gaunt interaction effects in boron isoelectronic sequence using coupled-cluster theory }
\author{\large Narendra Nath Dutta and \large  Sonjoy Majumder \\
 \small {\it Department of Physics and
Meteorology,\\ Indian Institute of Technology-Kharagpur, \\
\small   Kharagpur-721302, India \\
 }  }

\date{\today}

\begin{abstract}
\noindent In this paper, we have studied electron correlation and
Gaunt interaction effects in ionization potentials (IPs) and
hyperfine constants A of 2p$^2P_{1/2}$ and 2p$^2P_{3/2}$ states
along with the fine structure splitting (FSS) between them for
boron isoelectronic sequence using relativistic coupled-cluster
(RCC) method. The range of atomic number Z has been taken from 8
to 21. Gaunt contributions are presented at both Dirac-Fock (DF)
and coupled-cluster (CC) levels of calculations. The Gaunt
corrected correlated results of the IPs and the FSS are compared
with the results of NIST. Important correlation contributions like
core correlation, core polarisation, pair correlation  etc. are
studied for hyperfine constants A. Many distinct features of
correlation and relativistic effects are observed in these
studies. With best of our knowledge, within Gaunt limit, most of
the hyperfine constants are presented for the first time in the
literature.
\end{abstract}

\maketitle


\section{Introduction}

Researches in isoelectronic sequences of lighter atoms have been a
subject of  recent interest to study various atomic properties of
low-lying atomic states and the transitions between them
\cite{pachucki, yerokhin, hao}. Accurate estimations of these
properties require correlation corrections with Breit and quantum
electrodynamic (QED) effects \cite{jie}. However, individual
studies of all these effects are necessary to realize the relative
strengths between them with increasing atomic number for different
isoelectronic sequences. To implement these effects in different
many-body theories, suitable form of matrix elements of their
operators are necessary. From our literature survey, we have found
a number of such formulations for the Breit operator \cite{mann,
kim, pyper, hinze, grant}. However, in our work, we have
implemented the Gaunt interaction which is the magnetic part of
the Breit interaction \cite{grant} and is considered to be an
order of magnitude larger than the other part, called as the
retardation part \cite{mann}. Hence the Gaunt interaction is
considered to provide a useful approximation to the Breit
interaction \cite{mann}. The matrix element of the Gaunt operator
is reformulated to add with the coulomb operator in a self
consistent approach at both the DF as well as the CC level of
calculations.

In recent years, a number of theoretical calculations have been
performed on boron isoelectronic sequence which take into account
the Breit interaction in the atomic Hamiltonian \cite{hao, jie,
li, zhang, rancova, koc1, koc2, koc3, koc4, tachiev, safronova1,
safronova2, safronova3, merkelis, galavis, zheng}. These
calculations have been performed using different many-body
approaches like configuration-interaction technique \cite{hao,
rancova, galavis}, multi-configuration Dirac-Fock method
\cite{jie, li, zhang}, weakest bound electron potential model
theory \cite{zheng}, relativistic multireference
configuration-interaction technique \cite{koc1, koc2, koc3, koc4},
multiconfiguration Hartree-Fock method \cite{tachiev},
relativistic many-body perturbation theory \cite{safronova1,
safronova2, safronova3, merkelis} etc. Eliav et al. have
implemented Breit operator in the CC theory but treated only four
members of this sequence to calculate the IPs and the FSS
\cite{eliav}. Recently, the RCC calculations of this sequence have
been performed by Nataraj et al. \cite{nataraj} for Mg VIII, Si X
and S XII. They have performed the RCC calculations on different
transition properties among some low-lying states of these ions in
the basis of Dirac-Coulomb Hamiltonian. In their paper, they have
highlighted the requirement of the Breit interaction in the fine
structure splitting of 2$^2P$ term of this sequence.

Isoelectronic sequences are also very useful to study the trends
of relativistic and the different correlation effects in the
hyperfine properties of different ions \cite{panigrahy}. Panigrahy
et al. have investigated different correlation mechanisms on the
magnetic dipole hyperfine constants (A) of Li like systems by
many-body perturbation theory \cite{panigrahy}. As members of
boron isoelectronic sequence, hyperfine constants of C II, N III
and O IV have been calculated by J\"{o}nsson et al.
\cite{jonsson}. The QED and the interelectronic interaction
corrections in hyperfine properties have been analyzed by
Oreshkina et al. by large-scale configuration-interaction
Dirac-Fock-Sturm method for few members of this sequence
\cite{glazov}.

The purpose of this present paper is to analyze the correlation
and the Gaunt effects in the calculations of the IPs, the
hyperfine constants A of 2p$^2P_{1/2}$ and 2p$^2P_{3/2}$ states
and  the FSS between them for boron like systems using
relativistic coupled-cluster approach. The systematic
investigations of both these effects with increasing atomic number
can provide a comparative information about their contributions in
the calculations of these properties. Comparisons of the Gaunt
contributions at both the DF and the CC levels of calculations
have been explicitly studied to test the correlation effects on
these. Our final calculated results of the IPs and the FSS
including correlation and Gaunt effects, are compared with the
results of National Institute of Standards and Technology (NIST)
\cite{nist}. Graphical variations of these effects are shown with
increasing atomic number. The RCC method, applied in these
calculations, consist of single, double and partial triple
excitations \cite{dixit}. Different types of correlation effects
like core correlation, pair correlation and core polarisation in
the hyperfine constants are tabulated and are plotted to observe
their variations w.r.t. Z.

\section{Theory}

\subsection{Matrix element of Gaunt interaction operator}

The Breit interaction, introduced by Breit \cite{breit}, is the
first relativistic correction of the Coulomb interaction. The
frequency independent form of the Breit interaction between two
electrons, indicated by '1' and '2', is given by
\begin{equation}
H_B=-\frac{\overrightarrow{\alpha_1}\cdot\overrightarrow{\alpha_2}}{r_{12}}+
   \frac{1}{2}\left[\frac{\overrightarrow{\alpha_1}\cdot\overrightarrow{\alpha_2}}{r_{12}}-
   \frac{(\overrightarrow{\alpha_1}\cdot\overrightarrow{r_{12}})(\overrightarrow{\alpha_2}\cdot\overrightarrow{r_{12}})}{r^3_{12}}\right]
\end{equation}
where $\alpha_1$ and $\alpha_2$ are the corresponding Dirac
matrices and $r_{12}$ is the distance of separation between the
two electrons \cite{reither}. The overall Breit interaction is
contributed by magnetic part, called Gaunt interaction
\cite{gaunt} as stated earlier, represented by the first term of
Eq. 2.1 and the other part which includes the retardation effect,
called as retardation part, represented by the remaining part of
this equation.

Including Gaunt interaction with Coulomb interaction, the atomic
Hamiltonian of a {\it N} electron system is written in the form
\begin{equation}
H=\sum_{i=1}^{N}\left(c\overrightarrow{\alpha_i}\cdot\overrightarrow{p_i}+\left(\beta_i-1\right)
c^2 +V_{nuc}(r_i)+
\sum_{j<i}^{N}\left(\frac{1}{r_{ij}}-\frac{\overrightarrow{\alpha_i}\cdot\overrightarrow{\alpha_j}}{r_{ij}}\right)\right).
\end{equation}

The irreducible tensor operator form of the Gaunt interaction is
given by \cite{pyper,grant}
\begin{equation}
B_g=\sum_{\nu,L}(-1)^{\nu +L}
V_{\nu}(1,2)\left[\textbf{X}^{((1\nu)L)}(1).\textbf{X}^{((1\nu)L)}(2)\right]
\end{equation}
where $\nu=L-1$, $L$ or $L+1$ and
$\textbf{X}^{((1\nu)L)}(1)=\left[\bm{\alpha}_1\textbf{C}^{(\nu)}(1)\right]^{(L)}$
\cite{pyper}. In the long wavelength approximation \cite{pyper},
\begin{equation}
V_{\nu}(1,2)=\frac{r^{\nu}_{<}}{r^{\nu+1}_{>}}
\end{equation}
where $r_{<}/r_{>}=min/max(r_1, r_2)$.

The knowledge of general two electron matrix element of the Gaunt
operator is necessary to include this effect in the CC theory
which is derived from Ref. \cite{pyper} and is given as follows:
\begin{eqnarray}
\langle A_1B_2|B_{g}|C_1D_2\rangle&=&\langle
A_1B_2|\sum_{\nu,L}(-1)^{\nu +L}
V_{\nu}(1,2)\left[\textbf{X}^{((1\nu)L)}(1).\textbf{X}^{((1\nu)L)}(2)\right]|C_1D_2\rangle \nonumber \\
&=&\sum_{L,M}(-1)^{j_A-m_A+j_B-m_B+L-M}
 \left(%
\begin{array}{ccc}
  j_A & L & j_C \\
  -m_A& M & m_C \\
\end{array}%
\right)
\left(%
\begin{array}{ccc}
  j_B & L  & j_D \\
  -m_B& -M & m_D \\
\end{array}%
\right)\nonumber\\ &\times& X^{L}(ABCD).
\end{eqnarray}
Here operator strength, $X^{L}(ABCD)$  is written in the following
form:
\begin{eqnarray}
X^{L}(ABCD)&=&(-1)^{j_A+j_B+L+1}\sqrt{(2j_A+1)(2j_B+1)(2j_C+1)(2j_D+1)}
\left(%
\begin{array}{ccc}
  j_A & L & j_C \\
  \frac{1}{2}& 0 & -\frac{1}{2} \\
\end{array}%
\right)
\left(%
\begin{array}{ccc}
  j_B & L & j_D \\
  \frac{1}{2}& 0 & -\frac{1}{2} \\
\end{array}%
\right)\nonumber\\ &\times& \left[
\sum^{L+1}_{\nu=L-1}\Pi^{o}(\kappa_A,
\kappa_C,\nu)\Pi^{o}(\kappa_B,
\kappa_D,\nu)\sum^4_{\mu=1}r^{\nu}_{\mu}(ABCD)R^{\nu}_{\mu}(ABCD)\right].
\end{eqnarray}
 The factor
\begin{equation}
\Pi^{o}(\kappa_A, \kappa_C,\nu)=\frac{1}{2}\left[1+a_A
a_C(-1)^{j_A+j_{C}+\nu}\right]
\end{equation}
is associated with the parity selection rule of Gaunt interaction
operator which is opposite to the coulomb parity selection rule.
The values of $a_A$ and $a_C$ are +1 or $-$1 according to the
positive or negative kappa values, respectively. The coefficients
$r^{\nu}_{\mu}(ABCD)$ and the radial integrals
$R^{\nu}_{\mu}(ABCD)$ are presented in Table I and Table II,
respectively for values of $\mu$=1, 2, 3 or 4. For Table I, we
have
\begin{eqnarray}
P =
\begin{cases}
\dfrac{1}{L(2L-1)} \hspace*{3.1cm} for \hspace*{0.5cm} \nu=L-1\\
-\dfrac{(\kappa_A+\kappa_C)(\kappa_B+\kappa_D)}{L(L+1)} \hspace*{1cm} for \hspace*{0.5cm} \nu=L\\
\dfrac{1}{(L+1)(2L+3)} \hspace*{2.1cm} for \hspace*{0.5cm}
\nu=L+1,
\end{cases}
\end{eqnarray}
$\overline{k}=\kappa_C-\kappa_A$ and
$\overline{k'}=\kappa_D-\kappa_B$. $\dfrac{P_A(r)}{r}$ and
$\dfrac{Q_A(r)}{r}$ are the large and small components of the
radial part of the wavefunctions, respectively \cite{pyper}.

 At the Dirac-Fock level, we
need the knowledge of direct and exchange matrix elements of this
operator which are obtained by replacing $A=A$, $B=B$, $C=A$ and
$D=B$; and $A=A$, $B=B$, $C=B$ and $D=A$, respectively
\cite{grant}. However, the direct contribution to the Gaunt
interaction is zero \cite{mann}. So using the algebras of 3-$j$
symbols from Ref. \cite{grant}, we give the exchange matrix
element of this operator as follows:
\begin{equation}
\langle A_1B_2|B_g|B_1A_2\rangle =\sum_{L}(2j_B+1)\left(%
\begin{array}{ccc}
  j_A & L & j_B \\
  \frac{1}{2} & 0 & -\frac{1}{2} \\
\end{array}%
\right)^2\times\left[\sum^{L+1}_{\nu=L-1}\Pi^{o}(\kappa_A,
\kappa_B, \nu) \sum^4_{\mu=1}r^{\nu}_{\mu}(ABBA)
R^{\nu}_{\mu}(ABBA)\right].
\end{equation}
The coefficients $r^{\nu}_{\mu}(ABBA)$ are presented in Table III
and the radial integrals $R^{\nu}_{\mu}(ABBA)$ are obtained from
Table II by replacing $A=A$, $B=B$, $C=B$ and $D=A$. For Table
III, we have
\begin{eqnarray*}
P =
\begin{cases}
\dfrac{1}{L(2L-1)} \hspace*{2.5cm} for \hspace*{0.5cm} \nu=L-1\\
-\dfrac{(\kappa_A+\kappa_B)^2}{L(L+1)} \hspace*{2.0cm} for \hspace*{0.5cm} \nu=L\\
\dfrac{1}{(L+1)(2L+3)} \hspace*{1.5cm} for \hspace*{0.5cm} \nu=L+1
\end{cases}
\end{eqnarray*}
and $\overline{k}=-\overline{k'}=\kappa_B-\kappa_A$.

\clearpage

\begin{table}
\caption{Coefficients $r^{\nu}_{\mu}(ABCD)$}
\begin{tabular}{c c c c}
  \hline\hline
  & $\nu=L-1$  & $\nu=L$  & $\nu=L+1$ \\
  \hline
  $\mu=1$\hspace*{0.2cm} & $P(L+\overline{k})(L+\overline{k'})$ \hspace*{0.2cm}& $P$ & $P(\overline{k}-L-1)(\overline{k'}-L-1)$ \\
  $\mu=2$\hspace*{0.2cm} & $P(L-\overline{k})(L-\overline{k'})$ \hspace*{0.2cm}& $P$ & $P(\overline{k}+L+1)(\overline{k'}+L+1)$ \\
  $\mu=3$\hspace*{0.2cm} & $P(L+\overline{k})(\overline{k'}-L)$ \hspace*{0.2cm}& $P$ & $P(\overline{k}-L-1)(\overline{k'}+L+1)$ \\
  $\mu=4$\hspace*{0.2cm} & $P(\overline{k}-L)(L+\overline{k'})$ \hspace*{0.2cm}& $P$ & $P(\overline{k}+L+1)(\overline{k'}-L-1)$ \\
  \hline\hline
\end{tabular} \\
\label{tab:results1}
\end{table}

\begin{table}
\caption{Radial integrals $R^{\nu}_{\mu}(ABCD)$}
\begin{tabular}{c c}
\hline\hline
   & $\nu=L-1$, $L$ or $L+1$ \\
  \hline
  $\mu=1$ & \hspace*{0.5cm}
  $\int^{\infty}_0\int^{\infty}_0
P_A(r_1)Q_C(r_1)V_{\nu}(1,2)P_B(r_2)Q_D(r_2)dr_1dr_2$ \\
  $\mu=2$ &
  \hspace*{0.5cm}$\int^{\infty}_0\int^{\infty}_0
Q_A(r_1)P_C(r_1)V_{\nu}(1,2)Q_B(r_2)P_D(r_2)dr_1dr_2$ \\
 $\mu=3$  &
 \hspace*{0.5cm}$\int^{\infty}_0\int^{\infty}_0
P_A(r_1)Q_C(r_1)V_{\nu}(1,2)Q_B(r_2)P_D(r_2)dr_1dr_2$ \\
 $\mu=4$  &
 \hspace*{0.5cm}$\int^{\infty}_0\int^{\infty}_0
Q_A(r_1)P_C(r_1)V_{\nu}(1,2)P_B(r_2)Q_D(r_2)dr_1dr_2$  \\
  \hline\hline
\end{tabular}\\
\end{table}

\begin{table}
\caption{Coefficients $r^{\nu}_{\mu}(ABBA)$}
\begin{tabular}{c c c c}
  \hline\hline
  & $\nu=L-1$ & $\nu=L$ & $\nu=L+1$ \\
  \hline
  $\mu=1$ \hspace*{0.3cm} &  $-P(\overline{k}^2-L^2)$ \hspace*{0.3cm}& $P$ & $-P(\overline{k}^2-(L+1)^2)$ \\
  $\mu=2$ \hspace*{0.3cm} &  $-P(\overline{k}^2-L^2)$ \hspace*{0.3cm}& $P$ & $-P(\overline{k}^2-(L+1)^2)$ \\
  $\mu=3$ \hspace*{0.3cm} &  $-P(\overline{k}+L)^2$   \hspace*{0.3cm}& $P$ & $-P(\overline{k}-(L+1))^2 $ \\
  $\mu=4$ \hspace*{0.3cm} &  $-P(\overline{k}-L)^2$   \hspace*{0.3cm}& $P$ & $-P(\overline{k}+(L+1))^2 $ \\
  \hline\hline
\end{tabular}
\end{table}

\subsection{Coupled-Cluster theory}

The CC method is one of the most powerful highly correlated
many-body method due to its all order structure to account the
correlation effects \cite{lindgren, dutta}. This method is used
here for the one valence electron and has been described in
details elsewhere \cite{lindgren, mukherjee, Haque, Pal, ccsdt,
sahoo1, dixit}.

According to the CC theory, the correlated wavefunction of a
single valance atomic state having valance orbital '$v$' is
written in the form
\begin{equation}
|\Psi_v\rangle=e^{T}\{1+S_v\}|\Phi_v\rangle
\end{equation}
Here, $|\Phi_v\rangle$ is the corresponding DF state. $T$ is the
closed shell cluster operator which considers excitations from the
core orbitals and $S_v$ is the open shell cluster operator
corresponding to the valence electron '$v$' \cite{sahoo1}.

The  correlated expectation value of an operator $\hat{O}$ at any
particular atomic state $\Psi_v$ can be written as
\begin{eqnarray}\nonumber
O^{CC}_{vv}&=& \frac {\langle
\Psi_v|\hat{O}|\Psi_v\rangle}{\langle
\Psi_v|\Psi_v\rangle}\\
&=&\frac{\langle\Phi_v|\{1+S^{\dag}_v\}\overline{O}\{1+S_v\}|\Phi_v\rangle}{\langle\Phi_v|\{1+S^{\dag}_v\}e^{T^{\dag}}e^{T}\{1+S_v\}|\Phi_v\rangle}
\end{eqnarray}
where $\overline{O}=e^{T^{\dag}}Oe^{T}$.

\subsection{Hyperfine constant A }

The hyperfine constant A of a state represented by $|JM\rangle$ is
given by the following expression:
\begin{equation}
A=\mu_N g_I\frac{\langle
J||{\textbf{T}^{(1)}}||J\rangle}{\sqrt{J(J+1)(J+2)}}
\end{equation}
where $\mu_N$ is the nuclear magneton and $g_I$ is the g-factor of
the nucleus having nuclear spin I \cite{sahoo1,gopal}. The
operator $\textbf{T}^{(1)}$ and the single-particle reduced matrix
element of the electronic part of this operator is defined in Ref.
\cite{sahoo2, rajat}.

\section{Results and Discussions}

The CC calculations are based on the generation of DF orbitals.
Therefore, accurate descriptions of the radial part of the orbital
wavefunctions at the DF level is one of the building blocks for
accurate calculations. Here, these orbitals are considered to be
gaussian type orbitals (GTOs) and are generated in the environment
of $V^{N-1}$ potential of Dirac-Coulomb Hamiltonian where {\it N}
is the number of electrons of each single valance system
\cite{sahoo1}. The radial wavefunctions are generated on 750 grid
points which follow, $r_i=r_0\left[e^{h(i-1)}-1\right]$ with
$r_0=2\times10^{-6}$ and h=0.05. The nuclei are considered to obey
Fermi type distribution function \cite{sahoo1}. The GTOs are
obtained by using universal basis parameters: $\alpha_0$ and
$\beta$ \cite{sahoo1, gopal}. These parameters, presented in Table
IV, are optimized for each system with respect to the
wavefunctions obtained from GRASP 2 code where DF equations are
solved using numerical technique \cite{parpia}.

In the DF calculations, the number of bases are taken as 30, 25,
and 20 for s, p and d symmetries, respectively. However, in the CC
calculations, 12, 11 and 10 number of the bases are used including
all the bound orbitals for s, p and d symmetries, respectively.
These number of symmetries and bases are chosen in accordance with
the numerical convergence of the core correlation energies. In our
discussions, the DF results are calculated for Dirac-Coulomb
Hamiltonian. Also, correlation contributions ($\Delta^{corr}$) are
calculated by the differences between the CC and the DF results
for the Dirac-Coulomb Hamiltonian, i.e., Coulomb correlations. But
the Gaunt contributions ($\Delta^{Gaunt}$) are calculated by the
differences between the CC results for the Dirac-Coulomb-Gaunt
Hamiltonian and the Dirac-Coulomb Hamiltonian. The percentage
correlation contributions ($\%$ $\Delta^{corr}$) and Gaunt
contributions ($\%$ $\Delta^{Gaunt}$) are evaluated with respect
to the DF results and the CC results, respectively, for the
Dirac-Coulomb Hamiltonian.

In Table V, our calculated IPs at the DF level are presented along
with $\Delta^{corr}$ and $\Delta^{Gaunt}$. The final results which
are the sum of these three, i.e.,
DF+$\Delta^{corr}$+$\Delta^{Gaunt}$, are compared with the results
of NIST in the same table \cite{nist}. Except Ca XVI, the final
calculated IPs are in good agreement with the NIST results.
However, our results for the former element are within the
uncertainty limits (about $\pm$16000 cm$^{-1}$) of the
experimental measurements \cite {jie}. From Table V, one can see
$\Delta^{Gaunt}$ are negative everywhere and with increasing
atomic number their absolute values increase monotonically whereas
positive values of $\Delta^{corr}$ decrease and become negative at
$Z\ge17$ and $Z\ge18$ for 2p$^2P_{1/2}$ and 2p$^2P_{3/2}$ states,
respectively. According to Eliav et al, for Z=10, the estimated
correlation contributions for  2p$^2P_{1/2}$  and 2p$^2P_{3/2}$
states are 4806.93 and 4865.75 cm$^{-1}$, respectively whereas for
Z=18 these values are $-$1104.40 and $-$59.26 cm$^{-1}$,
respectively which are in well agreement with our calculations.
The IP of 2p$^2P_{3/2}$ state for Z=16 calculated at the DF level
is close to the total result. For this case, the absolute values
of $\Delta^{Gaunt}$ and $\Delta^{corr}$ are relatively close to
each other, but their signs are opposite. So the overall
contribution of these two effects do not change the DF result
significantly. However, the wavefunctions responsible for the DF
and final results are entirely different which is observed in the
calculations of the hyperfine properties as discussed later in the
present paper. As seen from the table, at higher Z values of the
sequence, $\Delta^{Gaunt}$ are comparable with $\Delta^{corr}$ in
the determinations of the IPs.

In Fig. 1 and Fig. 2, graphical variations of  $\%$
$\Delta^{corr}$ and $\%$ $\Delta^{Gaunt}$ to the IPs w.r.t. Z are
presented, respectively. The $\%$ $\Delta^{corr}$ first decrease
rapidly and then vary slowly whereas absolute values of the $\%$
$\Delta^{Gaunt}$ increase linearly with increasing Z. In Fig. 1,
the $\%$ $\Delta^{corr}$ curve of 2p$^2P_{1/2}$ states shows
slightly more fall compare to the curve of 2p$^2P_{3/2}$ states.
Even from Fig. 2, one can see the curve of 2p$^2P_{1/2}$ is
slightly more steep than that of 2p$^2P_{3/2}$. So from these two
figures, it is obvious that with increasing Z, both $\%$
$\Delta^{Gaunt}$ and $\%$ $\Delta^{corr}$ are more effective for
the former states than that of the latter states in determining
the IPs. The correlation and the Gaunt effects are found to vary
from +1$\%$ to $-$0.1$\%$ and $-$0.01$\%$ to $-$0.04$\%$,
respectively in the IPs.

In Table VI, the FSS between 2p$^2P_{1/2}$ and 2p$^2P_{3/2}$
states are tabulated with correlation and Gaunt contributions. The
final results are compared with the results of NIST \cite{nist}.
From this table one can find, except at Z=8, $\Delta^{corr}$ are
negative for rest of the Z and their absolute values increase with
increasing Z. But $\Delta^{Gaunt}$ are negative for all the Z and
their absolute values also increase with increasing Z. In the
cases of Mg VIII, Si X and S XII, our CC results
(DF+$\Delta^{corr}$) are in good agreement with the results of
Nataraj et al. \cite{nataraj}. However, the significant
improvement of the final results due to the inclusion of the Gaunt
interaction not only for these but also for the other ions are
noted from this table.

In Fig. 3, variation of percentage correlation and Gaunt
contributions to the FSS are plotted w.r.t. Z. This figure
highlights that from Z=9, absolute values of $\%$ $\Delta^{corr}$
first increase rapidly, then slow down and after Z=16, decrease
slowly whereas absolute values of $\%$ $\Delta^{Gaunt}$ decrease
systematically with increasing Z. Up to Z=9, $\%$ $\Delta^{Gaunt}$
dominate over $\%$ $\Delta^{corr}$, but after that the case is
reverse. From this figure. one can see, $\%$ $\Delta^{corr}$ vary
from +1.5$\%$ to $-$4.5$\%$ and $\%$ $\Delta^{Gaunt}$ vary from
$-$4.5$\%$ to $-$1.5$\%$ in the FSS. These show that both
$\Delta^{corr}$ and $\Delta^{Gaunt}$ are very much important in
accurate determinations of the FSS compare to the IPs.

In Table VII, the hyperfine constants A are tabulated with
correlation and Gaunt effects. In these calculations, the most
stable isotopes of the each elements are chosen and the $g_I$
values of these isotopes are calculated from Ref. \cite{raghavan}.
Here, we consider the magnitudes of the $g_I$ values neglecting
their signs and are presented in the same table. The
multiconfiguration Dirac-Hartree-Fock results within Breit-Pauli
approximation by J\"{o}nsson et al. for O IV (Z=8) of
2p$^2P_{1/2}$ and 2p$^2P_{3/2}$ states are 1647 and 324 MHz,
respectively which are in good agreement with our final results
\cite{jonsson}. This table clearly shows correlation contributions
arise as a dominating mechanism compared to the Gaunt
contributions in the determinations of the hyperfine constants.
Contrary to the IP, the DF result differ significantly from the
final result of 2p$^2P_{3/2}$ state for Z=16 due to the difference
of the wavefunctions between two different level of calculations.

 In Table VIII, important correlation contributions from core correlation
($\overline{O}-O$), pair correlation ($\overline{O}S_{1v}$+c.c.)
and the lowest order core polarisation ($\overline{O}S_{2v}$+c.c.)
along with correlations from the terms
$S^{\dag}_{2v}\overline{O}S_{2v}$ +c.c. and normalization
corrections to the hyperfine constants are reported \cite{gopal}.
Here c.c. stands for complex conjugate of the corresponding term
\cite{gopal}. The remaining correlation contributions come from
the terms like $S^{\dag}_{1v}\overline{O}S_{1v}$ +c.c. and
$S^{\dag}_{1v}\overline{O}S_{2v}$ +c.c. and the other effective
two-body terms \cite{gopal} which are not discussed here due to
their relatively small contributions. As seen from this table, the
pair correlation effects are positive, but core correlation and
core polarisation effects are opposite in sign between the fine
structure states. For lighter ions, considerable correlation
contributions are found to occur from the terms
$S^{\dag}_{2v}\overline{O}S_{2v}$ +c.c. for 2p$^2P_{3/2}$ states.

The variation of $\%$ $\Delta^{corr}$, i.e., the percentage of
total correlation contributions, w.r.t. Z to the hyperfine
constants are shown in Fig. 4 and Fig. 5 along with the different
correlation effects for 2p$^2P_{1/2}$ and 2p$^2P_{3/2}$ states,
respectively. The percentage correlation contributions of the
different correlation terms are calculated w.r.t. the DF results.
Like the IPs, here also $\%$ $\Delta^{corr}$ first decrease
rapidly and then decrease slowly with increasing Z. But unlike to
the IPs, $\%$ $\Delta^{corr}$ are positive everywhere. The $\%$
$\Delta^{corr}$ of 2p$^2P_{1/2}$ and 2p$^2P_{3/2}$ states vary
from +4.25$\%$ to +1.25$\%$ and +8$\%$ to +4$\%$, respectively. As
Z increase, absolute values of the percentage correlation
contributions from the different correlation terms decrease. Among
these, the core correlation effects are most stable with respect
to the other two correlation effects. At higher Z values, major
correlations come from the core correlations and the next higher
contributions come from the core polarisations.

In Table IX, Table X and Table XI, Gaunt contributions to the IPs,
the FSS and the hyperfine A constants are presented, respectively
at both the DF and the CC levels of calculations to show their
changes due to correlation effects. From Table IX, it is seen for
2p$^2P_{1/2}$ states where correlation effects on Gaunt
contributions are increasing systematically, for 2p$^2P_{3/2}$
states, it is increasing up to Z=15 and then start decreasing with
increasing Z. The same table also shows that correlation effects
on Gaunt contributions are relatively more stronger for
2p$^2P_{1/2}$ states compare to the 2p$^2P_{3/2}$ states. In the
IP's, correlation effects change the Gaunt contributions from the
DF to the CC levels by +7.50$\%$ to +2.94$\%$ and +3.73$\%$ to
$-$0.09$\%$ for 2p$^2P_{1/2}$ and 2p$^2P_{3/2}$ states,
respectively. However, in the FSS the changes are more stronger
which are about +17.34$\%$ to +15.13$\%$. As expected due to
relatively large correlation effects, the dramatic changes occur
in the hyperfine constants as seen from Table XI which almost
exhaust the Gaunt contributions at the DF levels and provide very
small to the CC levels. These changes are about +102.97$\%$ to
+88.60$\%$ and +88.89$\%$ to +75.61$\%$ for 2p$^2P_{1/2}$ and
2p$^2P_{3/2}$ states, respectively.

\clearpage

\begingroup
\squeezetable
\begin{table}
\scriptsize \caption{Universal basis parameters $\alpha_0$ and $\beta$}
\begin{tabular}{lrrrrrrrrrrrrrr}
\hline\hline
 Z & 8 & 9 & 10 & 11 & 12 & 13 & 14 & 15 & 16 & 17 & 18 & 19 & 20 & 21 \\
 \hline
 $\alpha_0$ & 0.00225 & 0.00275 & 0.00325 & 0.00350 & 0.00425 &
 0.00525 & 0.00625 & 0.00725 & 0.00825 & 0.00925 & 0.01025 &
 0.01125 & 0.01225 & 0.01325 \\
 $\beta$ & 2.73 & 2.73 & 2.73 & 2.73 & 2.73 & 2.73 & 2.73 & 2.73 &
 2.73 & 2.73 & 2.73 & 2.73 & 2.73 & 2.73 \\
\hline\hline
\end{tabular} \\
\label{tab:results1}
\end{table}
\endgroup

\begin{table*}
\caption{Calculated IPs  with correlation and Gaunt effects along
with the comparisons with the NIST results (in cm$^{-1}$).}
\begin{tabular}{lcrrrrrr}
\hline\hline
  Z & State & DF & $\Delta^{corr}$ & $\Delta^{Gaunt}$  &  Total$^a$ & NIST$^b$ \\
\hline
8 & 2p$^2P_{1/2}$ &    618204.38  &  5808.89 &   -73.72  &   623939.55 & 624382 \\
  & 2p$^2P_{3/2}$ &    617770.18  &  5803.55 &   -55.03  &   623518.70 & 623996 \\
9 & 2p$^2P_{1/2}$ &    915988.05  &  5307.30 &  -128.55  &   921166.80 & 921430\\
  & 2p$^2P_{3/2}$ &    915158.69  &  5324.07 &   -98.90  &   920383.86 & 920686\\
10& 2p$^2P_{1/2}$ &   1269209.36  &  4716.60 &  -204.62  &  1273721.34 & 1273820\\
  & 2p$^2P_{3/2}$ &   1267767.50  &  4764.47 &  -160.38  &  1272371.59 & 1272513\\
11& 2p$^2P_{1/2}$ &   1677809.33  &  4060.48 &  -305.45  &  1681564.36 & 1681700\\
  & 2p$^2P_{3/2}$ &   1675470.31  &  4152.24 &  -242.35  &  1679380.20 & 1679565\\
12& 2p$^2P_{1/2}$ &   2141817.07  &  3385.86 &  -433.87  &  2144769.06 & 2145100 \\
  & 2p$^2P_{3/2}$ &   2138220.31  &  3536.45 &  -347.56  &  2141409.20 & 2141798\\
13& 2p$^2P_{1/2}$ &   2661309.44  &  2677.04 &  -593.23  &  2663393.25 & 2662650\\
  & 2p$^2P_{3/2}$ &   2656009.14  &  2905.80 &  -478.73  &  2658436.21 & 2657760 \\
14& 2p$^2P_{1/2}$ &   3236356.19  &  1944.52 &  -787.29  &  3237513.42 & 3237300\\
  & 2p$^2P_{3/2}$ &   3228811.95  &  2274.68 &  -638.94  &  3230447.69 & 3230309\\
15& 2p$^2P_{1/2}$ &   3867061.50  &  1193.76 & -1019.32  &  3867235.94 & 3867100 \\
  & 2p$^2P_{3/2}$ &   3856629.52  &  1651.45 &  -830.90  &  3857450.07 & 3857401 \\
16& 2p$^2P_{1/2}$ &   4553546.98  &   411.06 & -1292.67  &  4552665.37 & 4552500\\
  & 2p$^2P_{3/2}$ &   4539468.86  &  1032.31 & -1057.46  &  4539443.71 & 4539365\\
17& 2p$^2P_{1/2}$ &   5295946.59  &  -371.31 & -1610.97  &  5293964.31 & 5293800\\
  & 2p$^2P_{3/2}$ &   5277342.47  &   444.22 & -1321.51  &  5276465.18 & 5276390\\
18& 2p$^2P_{1/2}$ &   6094410.81  & -1166.35 & -1977.55  &  6091266.91 & 6090500\\
  & 2p$^2P_{3/2}$ &   6070267.99  &  -118.69 & -1625.80  &  6068523.50 & 6067844\\
19& 2p$^2P_{1/2}$ &   6949103.86  & -1973.12 & -2395.89  &  6944734.85 & 6943800\\
  & 2p$^2P_{3/2}$ &   6918267.32  &  -652.16 & -1973.18  &  6915641.98 & 6914783\\
20& 2p$^2P_{1/2}$ &   7860203.67  & -2790.74 & -2869.39  &  7854543.54 & 7860000 \\
  & 2p$^2P_{3/2}$ &   7821366.32  & -1152.48 & -2366.49  &  7817847.35 & 7823480\\
21& 2p$^2P_{1/2}$ &   8827901.77  & -3618.45 & -3401.58  &  8820881.74 & 8820000 \\
  & 2p$^2P_{3/2}$ &   8779594.34  & -1616.56 & -2808.45  &  8775169.33 & 8774363\\
 \hline\hline
\end{tabular}

$^a$Total= DF+$\Delta^{corr}$+$\Delta^{Gaunt}$.\\

$^b$NIST $\rightarrow$ NIST results \cite{nist}.
\end{table*}

\begin{table}
\caption{Calculated FSS  between 2p$^2P_{1/2}$ and 2p$^2P_{3/2}$
states with correlation and Gaunt effects along with the
comparisons with the NIST results (in cm$^{-1}$) .}
\begin{tabular}{lcrrrrr}
\hline\hline
  Z& DF & $\Delta^{corr}$ & $\Delta^{Gaunt}$  &  Total$^a$ & NIST$^b$ \\
\hline
8  &   434.20  &     5.34  &  -18.69  &   420.85  &   386 \\
9  &   829.36  &   -16.77  &  -29.65  &   782.94  &   744  \\
10 &  1441.86  &   -47.87  &  -44.24  &  1349.75  &  1307 \\
11 &  2339.02  &   -91.76  &  -63.10  &  2184.16  &  2135 \\
12 &  3596.76  &  -150.59  &  -86.31  &  3359.86  &  3302 \\
13 &  5300.30  &  -228.76  & -114.50  &  4957.04  &  4890 \\
14 &  7544.24  &  -330.16  & -148.33  &  7065.75  &  6991 \\
15 & 10431.98  &  -457.69  & -188.42  &  9785.87  &  9699 \\
16 & 14078.12  &  -621.25  & -235.21  & 13221.66  &  13135 \\
17 & 18604.12  &  -815.53  & -289.46  & 17499.13  &  17410 \\
18 & 24142.82  & -1047.66  & -351.75  & 22743.41  &  22656  \\
19 & 30836.54  & -1320.96  & -422.71  & 29092.87  &  29017  \\
20 & 38837.35  & -1638.26  & -502.90  & 36696.19  &  36520 \\
21 & 48307.43  & -2001.89  & -593.13  & 45712.41  &  45637 \\
 \hline\hline
\end{tabular}

$^a$Total= DF+$\Delta^{corr}$+$\Delta^{Gaunt}$.\\

$^b$NIST $\rightarrow$ NIST results \cite{nist}.

\end{table}

\begin{table}
\caption{Calculated hyperfine constants A with Correlation and
Gaunt effects (in MHz).}
\begin{tabular}{lcrrrrr}
\hline\hline
 Z & $g_I$ & State & DF & $\Delta^{corr}$  & $\Delta^{Gaunt}$  &  Total$^a$  \\
\hline
8 & 0.7575 & 2p$^2P_{1/2}$ &  1595.69  &  64.98  &   0.03  &  1660.70 \\
  &        & 2p$^2P_{3/2}$ &   317.67  &  25.37  &  -0.02  &   343.02  \\
9 & 5.2577 & 2p$^2P_{1/2}$ & 18100.83  & 600.68  &  -0.10  & 18701.41  \\
  &        & 2p$^2P_{3/2}$ &  3598.48  & 239.80  &  -0.29  &  3837.99  \\
10& 0.4412 & 2p$^2P_{1/2}$ &  2310.26  &  64.83  &  -0.05  &  2375.04  \\
  &        & 2p$^2P_{3/2}$ &   458.56  &  26.97  &  -0.04  &   485.49  \\
11& 1.4784 & 2p$^2P_{1/2}$ & 11170.12  & 270.02  &  -0.49  & 11439.65  \\
  &        & 2p$^2P_{3/2}$ &  2213.27  & 117.65  &  -0.26  &  2330.66  \\
12& 0.3422 & 2p$^2P_{1/2}$ &  3582.42  &  77.29  &  -0.20  &  3659.51  \\
  &        & 2p$^2P_{3/2}$ &   708.45  &  35.37  &  -0.10  &   743.72  \\
13& 1.4566 & 2p$^2P_{1/2}$ & 20459.89  & 402.39  &  -1.42  & 20860.86  \\
  &        & 2p$^2P_{3/2}$ &  4037.56  & 191.52  &  -0.65  &  4228.43  \\
14& 1.1106 & 2p$^2P_{1/2}$ & 20390.77  & 368.23  &  -1.68  & 20757.32  \\
  &        & 2p$^2P_{3/2}$ &  4014.68  & 183.07  &  -0.75  &  4197.00 \\
15& 2.2632 & 2p$^2P_{1/2}$ & 53148.34  & 887.65  &  -5.15  & 54030.84  \\
  &        & 2p$^2P_{3/2}$ & 10438.26  & 461.21  &  -2.23  & 10897.24  \\
16& 0.4292 & 2p$^2P_{1/2}$ & 12658.08  & 196.88  &  -1.41  & 12853.55 \\
  &        & 2p$^2P_{3/2}$ &  2479.40  & 106.89  &  -0.60  &  2585.69  \\
17& 0.5479 & 2p$^2P_{1/2}$ & 19978.04  & 290.71  &  -2.54  & 20266.21 \\
  &        & 2p$^2P_{3/2}$ &  3902.05  & 164.37  &  -1.05  &  4065.37  \\
18& 0.3714 & 2p$^2P_{1/2}$ & 16518.00  & 225.90  &  -2.37  & 16741.53  \\
  &        & 2p$^2P_{3/2}$ &  3216.44  & 132.62  &  -0.96  &  3348.10  \\
19& 0.1433 & 2p$^2P_{1/2}$ &  7682.31  &  99.13  &  -1.23  &  7780.21  \\
  &        & 2p$^2P_{3/2}$ &  1491.10  &  60.23  &  -0.49  &  1550.84  \\
20& 0.3765 & 2p$^2P_{1/2}$ & 24077.93  & 294.13  &  -4.29  & 24367.77 \\
  &        & 2p$^2P_{3/2}$ &  4657.46  & 184.30  &  -1.68  &  4840.08 \\
21& 1.3590 & 2p$^2P_{1/2}$ &102723.90  &1191.39  & -20.18  &103895.11 \\
  &        & 2p$^2P_{3/2}$ & 19798.50  & 767.11  &  -7.75  & 20557.86 \\
 \hline\hline
\end{tabular}
\label{tab:results1}

$^a$Total= DF+$\Delta^{corr}$+$\Delta^{Gaunt}$.\\
\end{table}

\begin{table}
\caption{Important correlation contributions to the hyperfine
constants A (in MHz).}
\begin{tabular}{lcrrrrr}
\hline\hline
  Z & State & Core-Corr$^a$ & Pair-Corr$^b$ & Core-Polr$^c$ & $S^{\dag}_{2v}\overline{O}S_{2v}$ & Norm$^d$  \\
\hline
8 & 2p$^2P_{1/2}$&   -28.07  &   26.51  &   35.66  &   13.62 &   21.51 \\
  & 2p$^2P_{3/2}$&    11.04  &    5.25  &   -9.43  &   15.32 &    4.38 \\
9 & 2p$^2P_{1/2}$&  -311.09  &  215.09  &  341.76  &  119.27 &  270.91 \\
  & 2p$^2P_{3/2}$&   116.17  &   42.50  &  -87.33  &  123.50 &   54.41 \\
10& 2p$^2P_{1/2}$&   -38.98  &   20.66  &   37.65  &   12.10 &   36.84 \\
  & 2p$^2P_{3/2}$&    14.11  &    4.07  &   -9.41  &   11.76 &    7.32 \\
11& 2p$^2P_{1/2}$&  -185.61  &   76.62  &  158.66  &   47.50 &  185.94 \\
  & 2p$^2P_{3/2}$&    65.82  &   15.07  &  -40.17  &   43.75 &   36.46 \\
12& 2p$^2P_{1/2}$&   -58.75  &   19.82  &   45.66  &   12.67 &   61.28 \\
  & 2p$^2P_{3/2}$&    20.54  &    3.89  &  -11.31  &   11.24 &   11.84 \\
13& 2p$^2P_{1/2}$&  -331.92  &   93.45  &  238.57  &   61.09 &  357.17 \\
  & 2p$^2P_{3/2}$&   114.83  &   18.28  &  -58.14  &   52.56 &   67.84 \\
14& 2p$^2P_{1/2}$&  -327.95  &   78.19  &  218.16  &   52.14 &  361.01 \\
  & 2p$^2P_{3/2}$&   112.47  &   15.24  &  -52.42  &   43.65 &   67.27 \\
15& 2p$^2P_{1/2}$&  -849.01  &  173.54  &  523.73  &  117.76 &  951.09 \\
  & 2p$^2P_{3/2}$&   288.87  &   33.71  & -124.24  &   96.30 &  173.43 \\
16& 2p$^2P_{1/2}$&  -201.59  &   35.62  &  115.32  &   24.56 &  229.00 \\
  & 2p$^2P_{3/2}$&    68.05  &    6.89  &  -27.05  &   19.66 &   40.72 \\
17& 2p$^2P_{1/2}$&  -317.05  &   48.96  &  168.84  &   34.26 &  363.96 \\
  & 2p$^2P_{3/2}$&   106.13  &    9.44  &  -39.17  &   26.89 &   62.94 \\
18& 2p$^2P_{1/2}$&  -261.57  &   35.58  &  129.91  &   25.23 &  302.70 \\
  & 2p$^2P_{3/2}$&    86.75  &    6.83  &  -29.83  &   19.46 &   50.74 \\
19& 2p$^2P_{1/2}$&  -121.53  &   14.66  &   56.40  &   10.53 &  141.51 \\
  & 2p$^2P_{3/2}$&    39.89  &    2.80  &  -12.82  &    7.98 &   22.92 \\
20& 2p$^2P_{1/2}$&  -380.95  &   41.00  &  165.45  &   29.80 &  445.60 \\
  & 2p$^2P_{3/2}$&   123.56  &    7.80  &  -37.30  &   22.25 &   69.48 \\
21& 2p$^2P_{1/2}$& -1626.87  &  157.03  &  662.34  &  115.47 & 1909.14 \\
  & 2p$^2P_{3/2}$&   520.56  &   29.73  & -148.18  &   84.92 &  285.64 \\
 \hline\hline
\end{tabular}
\label{tab:results1}

$^a$Core-Corr $\longrightarrow$ Core correlation.\\
$^b$Pair-Corr $\longrightarrow$ Pair correlation. \\
$^c$Core-Polr $\longrightarrow$ Core polarisation.\\
$^d$Norm      $\longrightarrow$ Normalization correction.\\
\end{table}

\begin{table}
\caption{Gaunt contributions to the IPs at the DF and the CC
levels along with their differences, $\Delta{E_g}$ (in
cm$^{-1}$).}
\begin{tabular}{lcrrrrr}
\hline\hline
 Z  & State & DF & $\Delta{E_g}$  & CC    \\
\hline
8  & 2p$^2P_{1/2}$ &  -79.70  &  5.98  &  -73.72  \\
   & 2p$^2P_{3/2}$ &  -57.16  &  2.13  &  -55.03  \\
9  & 2p$^2P_{1/2}$ & -137.84  &  9.29  & -128.55  \\
   & 2p$^2P_{3/2}$ & -101.97  &  3.07  &  -98.90  \\
10 & 2p$^2P_{1/2}$ & -217.74  & 13.12  & -204.62  \\
   & 2p$^2P_{3/2}$ & -164.37  &  3.99  & -160.38  \\
11 & 2p$^2P_{1/2}$ & -323.04  & 17.59  & -305.45  \\
   & 2p$^2P_{3/2}$ & -247.17  &  4.82  & -242.35  \\
12 & 2p$^2P_{1/2}$ & -456.70  & 22.83  & -433.87  \\
   & 2p$^2P_{3/2}$ & -353.10  &  5.54  & -347.56  \\
13 & 2p$^2P_{1/2}$ & -622.14  & 28.91  & -593.23  \\
   & 2p$^2P_{3/2}$ & -484.92  &  6.19  & -478.73  \\
14 & 2p$^2P_{1/2}$ & -822.90  & 35.61  & -787.29  \\
   & 2p$^2P_{3/2}$ & -645.45  &  6.51  & -638.94  \\
15 & 2p$^2P_{1/2}$ &-1061.92  & 42.60  &-1019.32  \\
   & 2p$^2P_{3/2}$ & -837.42  &  6.52  & -830.90  \\
16 & 2p$^2P_{1/2}$ &-1344.01  & 51.34  &-1292.67  \\
   & 2p$^2P_{3/2}$ &-1063.64  &  6.18  &-1057.46  \\
17 & 2p$^2P_{1/2}$ &-1671.19  & 60.22  &-1610.97  \\
   & 2p$^2P_{3/2}$ &-1326.90  &  5.39  &-1321.51  \\
18 & 2p$^2P_{1/2}$ &-2047.36  & 69.81  &-1977.55  \\
   & 2p$^2P_{3/2}$ &-1629.95  &  4.15  &-1625.80  \\
19 & 2p$^2P_{1/2}$ &-2476.08  & 80.19  &-2395.89  \\
   & 2p$^2P_{3/2}$ &-1975.60  &  2.42  &-1973.18  \\
20 & 2p$^2P_{1/2}$ &-2960.70  & 91.31  &-2869.39  \\
   & 2p$^2P_{3/2}$ &-2366.61  &  0.12  &-2366.49  \\
21 & 2p$^2P_{1/2}$ &-3504.68  &103.10  &-3401.58  \\
   & 2p$^2P_{3/2}$ &-2805.77  & -2.68  &-2808.45  \\
 \hline\hline
\end{tabular}
\label{tab:results1}
\end{table}

\begin{table}
\caption{Gaunt contributions to the FSS  between 2p$^2P_{1/2}$ and
2p$^2P_{3/2}$ states at the DF and the CC levels along with their
differences, $\Delta{E_g}$ (in cm$^{-1}$) .}
\begin{tabular}{lcrrrrr}
\hline\hline
  Z& DF & $\Delta{E_g}$ & CC  \\
\hline
8  &   -22.54  &     3.85  &  -18.69  \\
9  &   -35.87  &     6.22  &  -29.65  \\
10 &   -53.37  &     9.13  &  -44.24  \\
11 &   -75.87  &    12.77  &  -63.10  \\
12 &  -103.60  &    17.29  &  -86.31  \\
13 &  -137.22  &    22.72  & -114.50  \\
14 &  -177.45  &    29.12  & -148.33  \\
15 &  -224.50  &    36.08  & -188.42  \\
16 &  -280.37  &    45.16  & -235.21  \\
17 &  -344.29  &    54.83  & -289.46  \\
18 &  -417.41  &    65.66  & -351.75  \\
19 &  -500.48  &    77.77  & -422.71  \\
20 &  -594.09  &    91.19  & -502.90  \\
21 &  -698.91  &   105.78  & -593.13 \\
 \hline\hline
\end{tabular}
\label{tab:results1}
\end{table}

\begin{table}
\caption{Gaunt contributions to the hyperfine constants A at the
DF and the CC levels along with their differences, $\Delta{A_g}$
(in MHz).}
\begin{tabular}{lcrrrrr}
\hline\hline
 Z & State & DF & $\Delta A_g$  & CC   \\
\hline
8 & 2p$^2P_{1/2}$ &  -1.01  &   1.04  &   0.03   \\
  & 2p$^2P_{3/2}$ &  -0.18  &   0.16  &  -0.02   \\
9 & 2p$^2P_{1/2}$ & -12.98  &  12.88  &  -0.10   \\
  & 2p$^2P_{3/2}$ &  -2.30  &   2.01  &  -0.29   \\
10& 2p$^2P_{1/2}$ &  -1.86  &   1.81  &  -0.05   \\
  & 2p$^2P_{3/2}$ &  -0.33  &   0.29  &  -0.04   \\
11& 2p$^2P_{1/2}$ &  -9.92  &   9.43  &  -0.49   \\
  & 2p$^2P_{3/2}$ &  -1.78  &   1.52  &  -0.26   \\
12& 2p$^2P_{1/2}$ &  -3.48  &   3.28  &  -0.20   \\
  & 2p$^2P_{3/2}$ &  -0.63  &   0.53  &  -0.10   \\
13& 2p$^2P_{1/2}$ & -21.64  &  20.22  &  -1.42   \\
  & 2p$^2P_{3/2}$ &  -3.90  &   3.25  &  -0.65   \\
14& 2p$^2P_{1/2}$ & -23.28  &  21.60  &  -1.68   \\
  & 2p$^2P_{3/2}$ &  -4.20  &   3.45  &  -0.75   \\
15& 2p$^2P_{1/2}$ & -65.15  &  60.00  &  -5.15   \\
  & 2p$^2P_{3/2}$ & -11.75  &   9.52  &  -2.23   \\
16& 2p$^2P_{1/2}$ & -16.57  &  15.16  &  -1.41   \\
  & 2p$^2P_{3/2}$ &  -2.99  &   2.39  &  -0.60   \\
17& 2p$^2P_{1/2}$ & -27.82  &  25.28  &  -2.54   \\
  & 2p$^2P_{3/2}$ &  -5.02  &   3.97  &  -1.05   \\
18& 2p$^2P_{1/2}$ & -24.38  &  22.01  &  -2.37   \\
  & 2p$^2P_{3/2}$ &  -4.39  &   3.43  &  -0.96   \\
19& 2p$^2P_{1/2}$ & -11.97  &  10.74  &  -1.23   \\
  & 2p$^2P_{3/2}$ &  -2.15  &   1.66  &  -0.49   \\
20& 2p$^2P_{1/2}$ & -39.52  &  35.23  &  -4.29   \\
  & 2p$^2P_{3/2}$ &  -7.10  &   5.42  &  -1.68   \\
21& 2p$^2P_{1/2}$ &-177.08  & 156.90  & -20.18   \\
  & 2p$^2P_{3/2}$ & -31.78  &  24.03  &  -7.75   \\
 \hline\hline
\end{tabular}
\label{tab:results1}

\end{table}

\clearpage

\begin{figure}
\begin{center}
\includegraphics[width=0.8\textwidth]{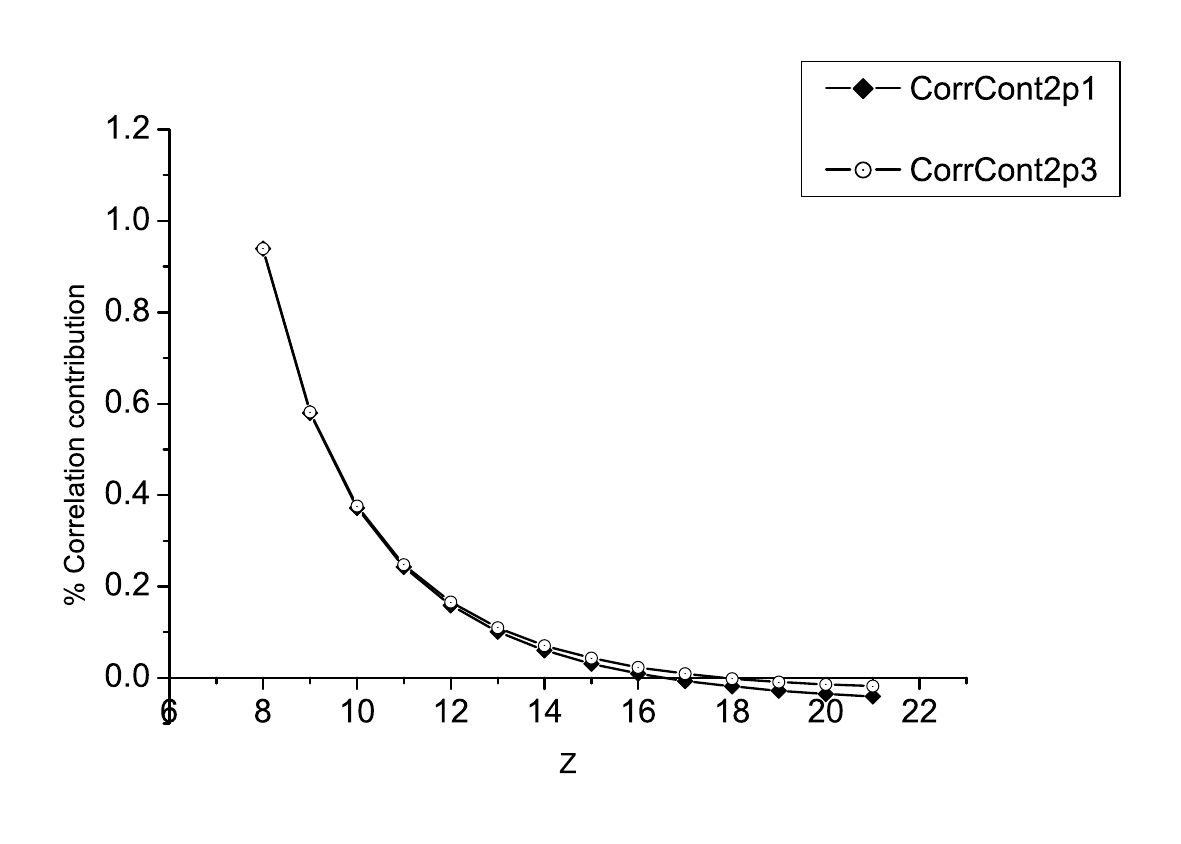}
\caption{\label{label} Percentage correlation contributions
(CorrCont) to the IPs of 2p$^2P_{1/2}$ (2p1) and 2p$^2P_{3/2}$
(2p3) states.} \label{fig1}
\end{center}
\end{figure}

\begin{figure}
\begin{center}
\includegraphics[width=0.8\textwidth]{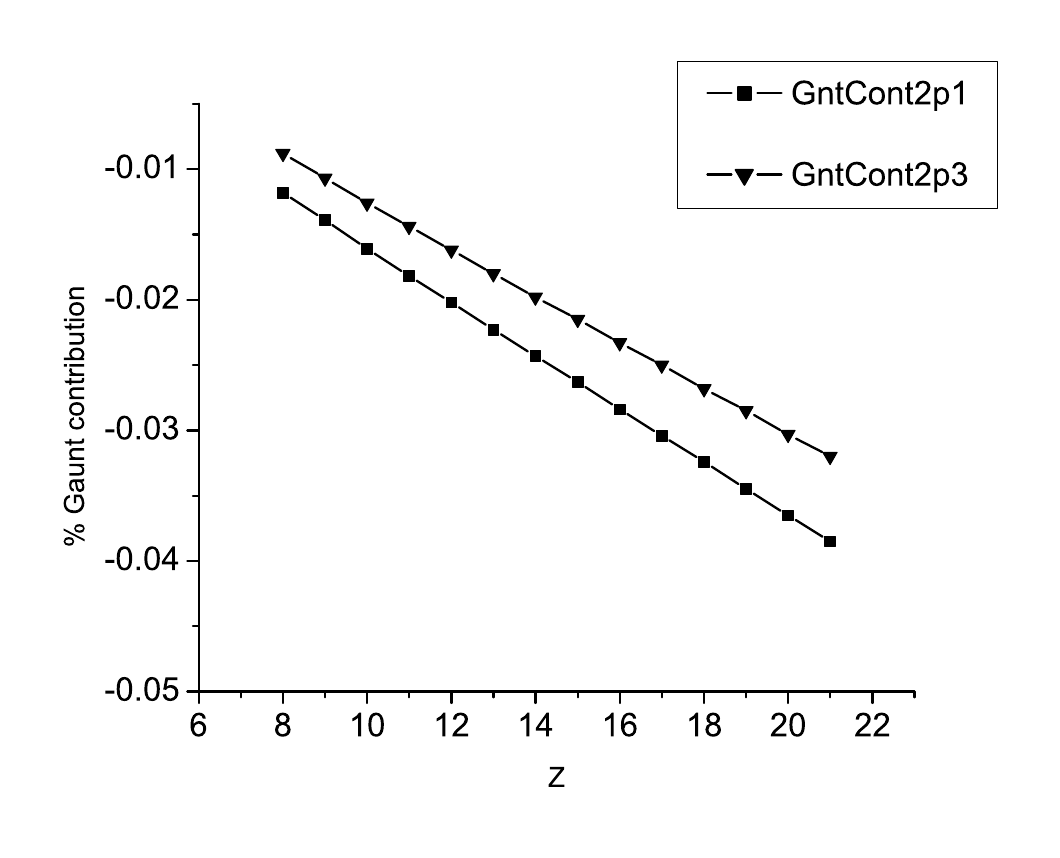}
\caption{\label{label}Percentage Gaunt contributions (GntCont) to
the IPs of 2p$^2P_{1/2}$ (2p1) and 2p$^2P_{3/2}$ (2p3) states.}
\label{fig2}
\end{center}
\end{figure}

\begin{figure}
\begin{center}
\includegraphics[width=0.8\textwidth]{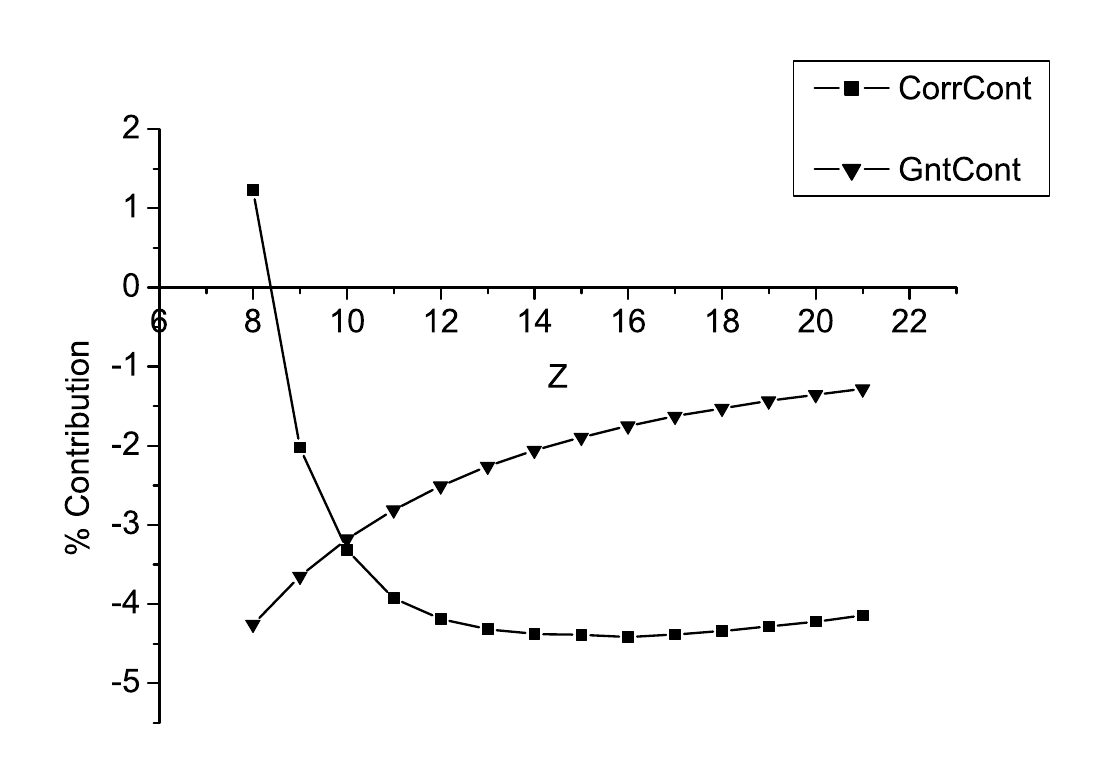}
\caption{\label{label}Percentage correlation (CorrCont) and Gaunt
contributions (GntCont) to the FSS between 2p$^2P_{1/2}$ and
2p$^2P_{3/2}$ states.} \label{fig3}
\end{center}
\end{figure}

\begin{figure}
\begin{center}
\includegraphics[width=0.8\textwidth]{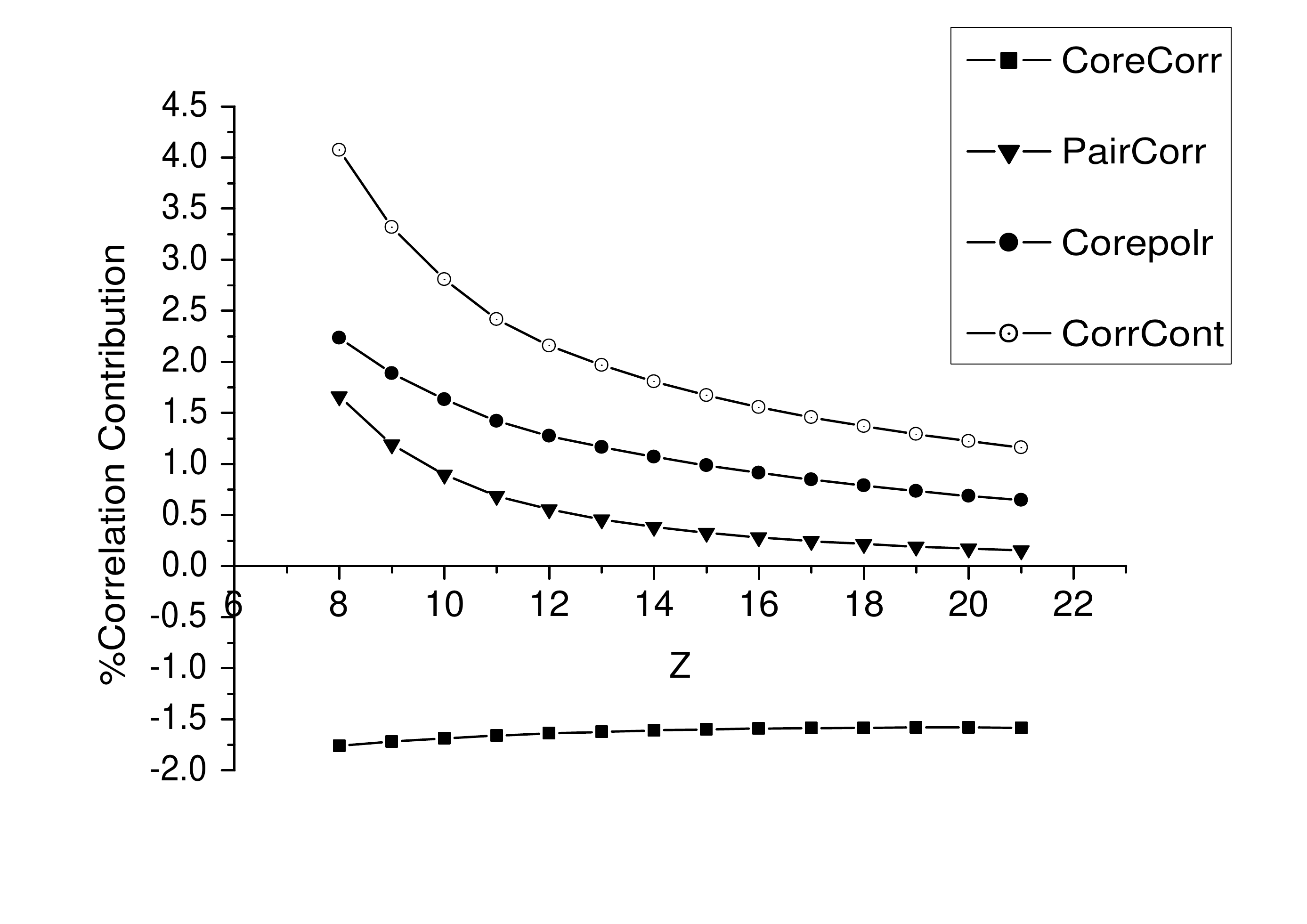}
\caption{\label{label} Percentage of total correlation
contributions (TotCorrCont) with core correlation (CoreCorr), pair
correlation (PairCorr) and core polarisation (CorePolr) effects in
the hyperfine constants A of 2p$^2P_{1/2}$ states.} \label{fig5}
\end{center}
\end{figure}

\begin{figure}
\begin{center}
\includegraphics[width=0.8\textwidth]{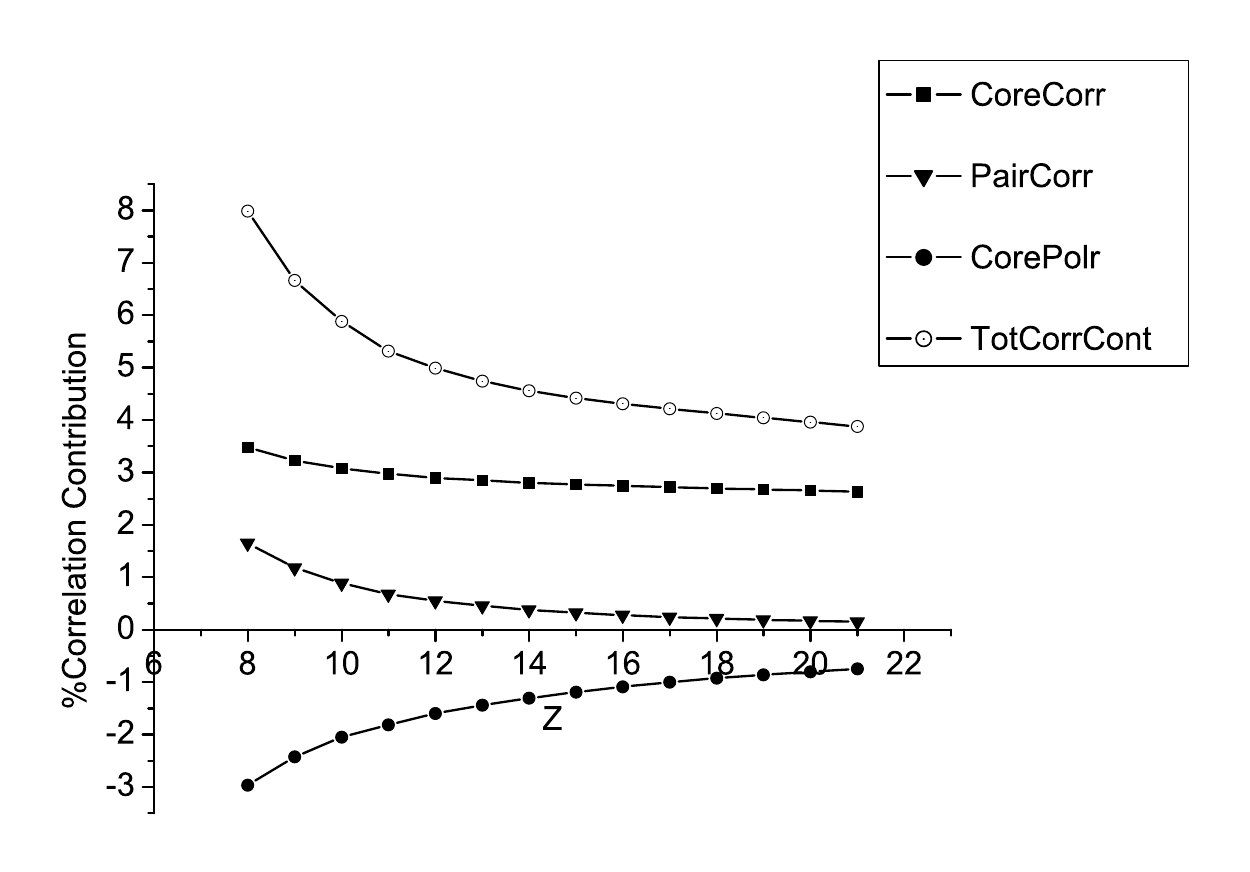}
\caption{\label{label} Percentage of total correlation
contributions (TotCorrCont) with core correlation (CoreCorr), pair
correlation (PairCorr) and core polarisation (CorePolr) effects in
the hyperfine constants A of 2p$^2P_{3/2}$ states. } \label{fig6}
\end{center}
\end{figure}

\clearpage

\section{Conclusion}

Detail analysis of the electron correlation and the Gaunt
contributions to the IPs and the hyperfine constants A have been
performed for first two low-lying states of boron like systems
using the RCC approach. We have reported the important role of
these two effects in the determinations of the FSS. Gaunt
contributions from the DF to the CC levels of calculations have
been discussed elaborately. The strengths of the correlation and
the Gaunt effects among all these properties with increasing Z
have been established. In the framework of the RCC theory,
contributions from the different correlation terms to the
hyperfine constants A have been studied descriptively. We hope, in
future, our study will be extended to incorporate the retardation
as well as the QED effects in both the DF and the CC levels of
calculations for more accurate descriptions of all these
properties. This will be also useful to judge the relative
strengths of all these effects with increasing Z not only for this
sequence but also for all the other isoelectronic sequences having
higher degree of correlation.

\begin{acknowledgments}
We are grateful to Prof B P Das and Dr R K Chaudhuri , Indian
Institute of Astrophysics, Bangalore, India and Dr B K Sahoo,
Physical Research Laboratory, Ahmedabad, India for providing the
CC and the Hyperfine code in which we have implemented the Gaunt
interaction. We are very much thankful of Dr A D K Singh and Mr B
K Mani, Physical Research Laboratory, Ahmedabad, India for
valuable suggestions of implementing the Gaunt interaction in the
CC code. We appreciate the help from Dr G Dixit, Center for
Free-Electron Laser Science, Hamburg, Germany. We would like to
recognize the support of Council of Scientific and Industrial
Research (CSIR), India for funding.
\end{acknowledgments}

\end{document}